\documentclass[conference]{IEEEtran}
\IEEEoverridecommandlockouts
\usepackage{cite}
\usepackage{amsmath,amssymb,amsfonts,texlogos}
\usepackage{algorithmic}
\usepackage{graphicx}
\usepackage{textcomp}
\usepackage{multicol,lipsum}
\usepackage{mathtools, cuted}
\usepackage{xcolor}
\def\BibTeX{{\rm B\kern-.05em{\sc i\kern-.025em b}\kern-.08em
    T\kern-.1667em\lower.7ex\hbox{E}\kern-.125emX}}

 \usepackage[linesnumbered,ruled]{algorithm2e}
\usepackage{amssymb,amsthm,textcomp}

\usepackage{url}
\usepackage[final]{pdfpages}
\usepackage{listings}
\usepackage{textcomp}
\usepackage{color} 
\usepackage{comment}

\usepackage{amsthm}
\usepackage{graphicx} 
\usepackage{mathtools}

\DeclareMathOperator*{\argmin}{arg\,min}
\usepackage{MnSymbol} 
\usepackage{multirow}
\usepackage{color}
\usepackage{dblfloatfix}
\usepackage{xifthen}
\newcommand{\oneline}[1]{%
  \newdimen{\namewidth}%
  \setlength{\namewidth}{\widthof{#1}}%
  \ifthenelse{\lengthtest{\namewidth < \textwidth}}%
  {#1}
  {\resizebox{\textwidth}{!}{#1}}
}
 \usepackage[protrusion=true,expansion=true]{microtype}
\usepackage[section]{placeins}
\usepackage{lipsum}
\usepackage{paralist}
\usepackage{wrapfig}
\theoremstyle{remark}
\usepackage{tabularx}
\newtheorem{definition}{Definition}[]

\newtheorem{proposition}{Proposition}[]
\newtheorem{lemma}{Lemma}[]

\newtheorem{corollary}{Corollary}[]
\usepackage[font=footnotesize]{caption}
\usepackage{lipsum,graphicx,subcaption}
\makeatletter
\newcommand\footnoteref[1]{\protected@xdef\@thefnmark{\ref{#1}}\@footnotemark}
\makeatother
\usepackage[inner=0.675in,outer=0.675in,top=0.7in,bottom=1in,pdftex]{geometry}
\usepackage[protrusion=true,expansion=true]{microtype}
\usepackage[font=footnotesize]{caption}
\begin{document}
\title{{Optimal Jammer Placement in UAV-assisted Relay Networks}}

\author{\IEEEauthorblockN{Seyyedali Hosseinalipour}
\IEEEauthorblockA{\textit{ECE Department} \\
\textit{North Carolina State University}\\
Raleigh, USA \\
shossei3@ncsu.edu}
\and
\IEEEauthorblockN{Ali Rahmati}
\IEEEauthorblockA{\textit{ECE Department} \\
\textit{North Carolina State University}\\
Raleigh, USA \\
arahmat@ncsu.edu}
\and
\IEEEauthorblockN{Huaiyu Dai}
\IEEEauthorblockA{\textit{ECE Department} \\
\textit{North Carolina State University}\\
Raleigh, USA \\
hdai@ncsu.edu}
}
\maketitle
\begin{abstract}
We consider the relaying application of unmanned aerial vehicles (UAVs), in which UAVs are placed between two transceivers (TRs) to increase the throughput of the system. Instead of studying the placement of UAVs as pursued in existing literature, we focus on investigating the placement of a \textit{jammer} or \textit{a major source of interference} on the ground to effectively degrade the performance of the system, which is measured by the maximum achievable data rate of transmission between the TRs. We demonstrate that the optimal placement of the jammer is in general a non-convex optimization problem, for which obtaining the solution directly is intractable. Afterward, using the inherent characteristics of the signal-to-interference ratio (SIR) expressions, we propose a tractable approach to find the optimal position of the jammer. Based on the proposed approach, we investigate the optimal positioning of the jammer in both \textit{dual-hop} and \textit{multi-hop} UAV relaying settings. Numerical simulations are provided to evaluate the performance of our proposed method.
\end{abstract}
 \vspace{-4mm}
\section{Introduction}\label{sec:intro}
\noindent Applications of unmanned aerial vehicles (UAVs) in wireless communication has attracted lots of attentions due to their ease of deployment and 3D movement capability, where one of their recent applications is data relaying~\cite{8424236}. On the other hand, jamming can degrade the performance of relay networks, which should be carefully addressed in practice. Although jamming and anti-jamming approaches have been  investigated
in wireless networks \cite{feng2014jammer,feng2017fast, allouche2017secure}, in the context of UAV-assisted networks, the current state of the art still lacks maturity \cite{krishna2017review}.

Optimal jammer placement has been studied in the context of network partitioning in wireless networks, e.g.,
 \cite{feng2014jammer}, \cite{feng2017fast}. In these works, the authors  investigate the effective jammer positioning
to partition a wireless network into multiple residual sub-networks with 
a constraint on the number of jammers.
 It is shown that there is a trade-off between the number of required jammers
 and the maximum order, i.e., the number of functional nodes, of the residual sub-networks.
Another application of jamming is providing a secure communication for sensitive information, where the usage of friendly
jammers to protect  sensitive communications is common \cite{allouche2017secure,arkin2015optimal}.  In
\cite{allouche2017secure}, the placement and power consumption of jammers is optimized in space and time
 to guarantee information-theoretic security for a secure communication. The aim is to prevent the  eavesdroppers outside the  protected zone from having a knowledge about the transmitted data.
A similar problem is studied in
\cite{arkin2015optimal}, where jamming via transmitting artificial noise is considered to protect
the communication from eavesdroppers. More discussions on (anti-)jamming techniques can be found in~\cite{grover2014jamming}.
Moreover, there is a body of literature devoted to \textit{jammer localization}, i.e., detecting the location of jammers, in wireless networks~\cite{7750577}. 

In the context of UAV relay networks, we were among the first to study the placement optimization and trajectory design for UAV relays to evade the interference caused by the jammers \cite{hosseinalipour2019interference,hosseinalipour2019interference2,rahmati2019interference,rahmati2019dynamic22}. Considering a major source of interference (MSI), the optimal placement of the UAV relays along with identifying the minimum number of required UAVs to satisfy a desired communication quality are studied in \cite{hosseinalipour2019interference,hosseinalipour2019interference2}. A joint power allocation and trajectory design is proposed in \cite{rahmati2019interference,rahmati2019dynamic22} to evade the interference caused by another established wireless network.
In \cite{hu2019proactive},  the optimal position and jamming power of a legitimate UAV monitor are obtained to maximize the average surveillance rate. In \cite{zhong2018secure}, a scenario is studied where a UAV transmits artificial noise to confuse
the ground eavesdropper for protection of the transmitted data.
In \cite{wang2018trajectory},  an anti-jamming approach is proposed in which the UAVs dynamically
adjust their trajectory.
Nevertheless, none of the aforementioned works investigates efficient degradation of the communication quality of UAV relay networks from the perspective of a jammer, which is our main motivation. 

In this paper, we consider a terrestrial jammer or MSI that aims to effectively deteriorate the communication quality of a UAV-assisted relay network working in the decode-and-forward relaying mode. We consider a two-way communication scenario, where the UAVs function as two-way relays between two terrestrial transceivers. The goal is to obtain the optimal placement of the terrestrial jammer to minimize the maximum achievable data rate of transmission between the terrestrial transceivers. We note that the optimal jammer placement problem belongs to the family of non-convex optimization problems, for which direct derivation of the solution is in general intractable. Using the inherent characteristics of the signal-to-interference ratio (SIR) expressions that result in piece-wise convexity of the objective function, we propose two efficient algorithms with polynomial complexity to obtain the optimal position of the jammer in the dual-hop and multi-hop UAV relay networks. Numerical simulations are conducted to reveal the performance of our proposed approach.
\section{Preliminaries and Problem Formulation}\label{pre}
\vspace{-1mm}
\subsection{Preliminaries}
\noindent We consider a two-way communication between a pair of transceivers (TRs), named TR\_1 and TR\_2, both engaged in transmitting and receiving the data. We assume a \textit{left-handed coordination system} $(x,y,h)$, and, without loss of generality, TR\_1 and TR\_2 are assumed to be located at $(0,0,0)$ and $(D,0,0)$, respectively. To improve the quality of communication, a set of UAV relays are placed between the TRs. We aim to effectively place a jammer/MSI on the ground to maximally deteriorate the communication performance of the system. Let $(x_{_{\textrm{MSI}}},y_{_{\textrm{MSI}}},{h}_{_{\textrm{MSI}}}=0)$ denote the position of the MSI.\footnote{Considering the MSI to be a flying UAV with a fixed altitude ${h}_{_{\textrm{MSI}}}=\hat{h}_{_{\textrm{MSI}}}$ incurs minor modifications in the derivations.}  The transmission powers of TR\_1, TR\_2, and the MSI are denoted by $p_{_{\textrm{TR\_1}}}$, $p_{_{\textrm{TR\_2}}}$, and $p_{_{\textrm{MSI}}}$, respectively.
 We consider both the line-of-sight (LoS) and the non-line-of-sight (NLoS) channel models, for which the path-loss is given by~\cite{mozaffari2017mobile}:
  \vspace{-1.5mm}
\begin{equation}
    L^{\textrm{LoS}}_{i,j}= \mu_{_{\textrm{LoS}}} d_{i,j}^{\alpha},\;\;L^{\textrm{NLoS}}_{i,j}=\mu_{_{\textrm{NLoS}}} d_{i,j}^{\alpha},
      \vspace{-1.0mm}
\end{equation}
where $\mu_{_{\textrm{LoS}}}\triangleq C_{_{\textrm{LoS}}}\left(4\pi f_c/c\right)^\alpha$, $\mu_{_{\textrm{NLoS}}}\triangleq C_{_{\textrm{NLoS}}}\left(4\pi f_c/c\right)^\alpha$, $C_{_{\textrm{LoS}}}$ ($C_{_{\textrm{NLoS}}}$) is the excessive path loss factor incurred by shadowing, scattering, etc., in the LoS (NLoS) link, $f_c$ is the carrier frequency, $c$ is the speed of light, $\alpha=2$ is the path-loss exponent, and $d_{i,j}$ is the Euclidean distance between node $i$ and node $j$. The link between two UAVs (air-to-air) is modeled using the LoS model, while the link between the MSI and a TR (ground-to-ground) is modeled based on the NLoS model. For the link between a UAV and the TRs or the MSI (air-to-ground and ground-to-air), we denote the path loss between a UAV $i$ and terrestrial node $j$ by $\eta_{_{\textrm{NLoS}}}d_{ij}^{2}$ (for more information on $\eta_{_{\textrm{NLoS}}}$ please refer to~\cite{hosseinalipour2019interference2} and references therein). Due to the geographical limitations, direct communication between the TRs is not considered. While the above channel models are relatively simple, they represent the current art in UAV modeling, and facilitate the derivation of many interesting results in current literature, e.g.,~\cite{8424236,mozaffari2017mobile}.
\vspace{-1mm}
\subsection{Problem Formulation}\label{sec:bigPicture}
\begin{figure}[t]
\includegraphics[width=8.8cm,height=2.1cm]{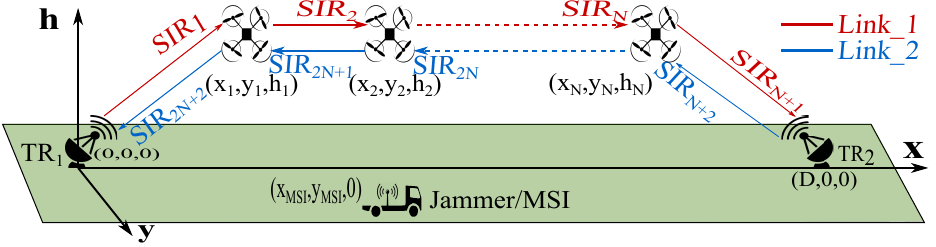}
		\caption{A jammer/MSI that aims to deteriorate the communication performance in multi-hop UAV relay setting.}
		 \label{fig:multiple}
		  \vspace{-1.3mm}
		 \end{figure}

As shown in Fig.~\ref{fig:multiple}, consider $N$ UAVs between the TRs, where the location of $\textrm{UAV}_i$ is denoted by $(x_i,y_i,z_i)$.  Let us define Link\_1 as the transmission link from TR\_1 to TR\_2 (when TR\_1 acts as a transmitter and TR\_2 acts as a receiver), and Link\_2 as the transmission link from TR\_2 to TR\_1. It is assumed that the UAVs utilize the same frequency but  different time slots to avoid mutual interference among the UAVs. We consider \textit{an interference limited environment}, where the power of noise is negligible compared to that  of interference caused by the MSI, and thus the SIR is used to describe the quality of communication. For Link\_1, let $\textrm{SIR}_i$ denote the SIR at $\textrm{UAV}_i$, $1\leq i\leq N$, and $\textrm{SIR}_{N+1}$ denote the SIR at TR\_2. Similarly, for Link\_2, $\textrm{SIR}_{N+2+i}$ denotes the SIR at $\textrm{UAV}_{N-i}$, $0\leq i\leq N-1$, and $\textrm{SIR}_{2N+2}$ denotes the SIR at TR\_1. Assuming decode-and-forward relaying, the SIR of Link\_1 and Link\_2 are given by:

 {\small
\begin{equation}\label{SIR_L_1_multi}\hspace{-13.9mm}
 \textrm{SIR}_{\textrm{Link\_1}}(x_{_{\textrm{MSI}}},{y}_{_{\textrm{MSI}}})=\min\{\textrm{SIR}_{i}(x_{_{\textrm{MSI}}},{y}_{_{\textrm{MSI}}}): 1\leq i\leq N+1\},
\hspace{-10mm}
\end{equation}
\begin{equation}\label{SIR_L_2_multi}\hspace{-13.9mm}
    \textrm{SIR}_{\textrm{Link\_2}}(x_{_{\textrm{MSI}}},{y}_{_{\textrm{MSI}}})=\min\{\textrm{SIR}_{N+i+2}(x_{_{\textrm{MSI}}},{y}_{_{\textrm{MSI}}}): 0\leq i\leq N\}.
    \hspace{-10mm}
\end{equation}
}
The goal of the jammer is to locate itself to effectively degrade the \textit{maximum achievable data rate of transmission between the TRs}. Assuming the same bandwidth for both links, this is equivalent to minimizing the maximum value of the $\textrm{SIR}$ of the links denoted by $\textrm{SIR}_{\textrm{max}}= \max (\textrm{SIR}_{\textrm{Link\_1}},\textrm{SIR}_{\textrm{Link\_2}})$. Thus,
\begin{equation}\label{MainProb22}
\begin{aligned}
    (x^*_{_{\textrm{MSI}}},y^*_{_{\textrm{MSI}}})=\argmin_{ x_{_{\textrm{MSI}}},y_{_{\textrm{MSI}}}}\hspace{2mm}\max \{\hspace{-.4mm}&\textrm{SIR}_{\textrm{Link\_1}}(\hspace{-.4mm}x_{_{\textrm{MSI}}},{y}_{_{\textrm{MSI}}}\hspace{-.4mm}),\\
    &\hspace{0.4mm}\textrm{SIR}_{\textrm{Link\_2}}(\hspace{-.4mm}x_{_{\textrm{MSI}}},{y}_{_{\textrm{MSI}}}\hspace{-.4mm}) \hspace{-.4mm}\}.
    \end{aligned}
\end{equation} 
 The SIR expressions are convex functions with respect to (w.r.t) $x_{_{\textrm{MSI}}}$ and $y_{_{\textrm{MSI}}}$ (see  \eqref{eq:SIRsLtR}). However, since the minimum of convex functions is not necessary convex, $\textrm{SIR}_{\textrm{Link\_1}}$ and $\textrm{SIR}_{\textrm{Link\_2}}$ are, in general, non-convex functions w.r.t the position of the jammer. This results in non-convexity of our main problem in~\eqref{MainProb22}, which makes  classic convex optimization techniques irrelevant and obtaining the solution non-trivial. In this work, we aim to develop a tractable analytical approach to solve this problem. Given the fact that tackling the problem where $x_{_{\textrm{MSI}}},y_{_{\textrm{MSI}}}$ are both variable is highly nontrivial, we fix one of those coordinates, which is $y_{_{\textrm{MSI}}}$ in this work such that $y_{_{\textrm{MSI}}}=\hat{y}_{_{\textrm{MSI}}}$, and obtain $x^*_{_{\textrm{MSI}}}$. Even with this assumption, the problem remains to be non-convex and non-trivial. Given the notable low complexity of our proposed method, one can obtain $x^*_{_{\textrm{MSI}}}$ for a set of given $y_{_{\textrm{MSI}}}$-s and choose the best solution among them. Also, the proposed methodology can be easily applied to the case where $x_{_{\textrm{MSI}}}$ is fixed and $y_{_{\textrm{MSI}}}$ is variable. Thus, one can set $x_{_{\textrm{MSI}}}=x^*_{_{\textrm{MSI}}}$ to obtain the corresponding $y^*_{_{\textrm{MSI}}}$ in a successive manner. Throughout, we assume that the MSI is mounted on a vehicle with the feasible moving area confined by $-x^-_{jam}\leq x_{_{\textrm{MSI}}}\leq x^+_{jam}$, where $x^+_{jam}\geq D,~-x^-_{jam}\leq 0$. To facilitate the discussion, we first investigate the problem in the dual-hop setting, which itself is of interest, and then extend the study to the multi-hop setting.

\vspace{-1mm}
\section{Jammer Placement in Dual-hop Setting}\label{sec:singleUAV}
\begin{figure}[t]
\includegraphics[width=8.9cm,height=2.3cm]{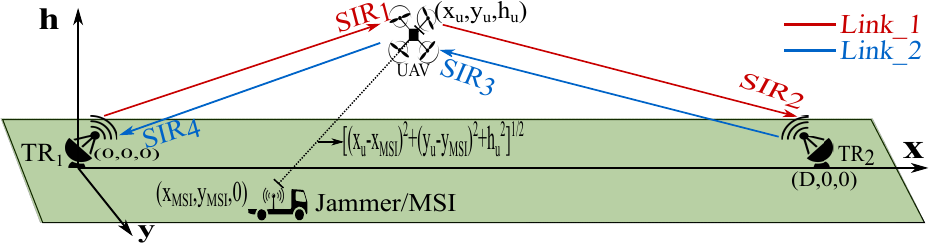}
		\caption{A jammer/MSI that aims to deteriorate the communication performance in dual-hop UAV relay setting.}
		 \label{fig:single}
		 \vspace{-1mm}
		 \end{figure}

\noindent Consider the jammer placement in the dual-hop setting, where the data is relayed via a single UAV located at $(x_u,y_u,h_u)$ with transmission power $p_u$ (see Fig.~\ref{fig:single}). The  SIR expressions are given by:
\vspace{-2mm}
\begin{equation}\label{SIR1}
{\small
\begin{aligned}
    \textrm{SIR}_1(x_{_{\textrm{MSI}}},\hat{y}_{_{\textrm{MSI}}})\hspace{-1mm}=\hspace{-1mm}\frac{p_{_{\textrm{TR\_1}}} \hspace{-.12mm} \left(\hspace{-.15mm}\left(\hspace{-.15mm}x_u-x_{_{\textrm{MSI}}}\hspace{-.15mm}\right)^2\hspace{-.15mm}+\hspace{-.14mm}\left(\hat{y}_{_{\textrm{MSI}}}-y_u\right)^2\hspace{-.15mm}+\hspace{-.12mm}h_u^2\hspace{-.15mm}\right)}{p_{_{\textrm{MSI}}}\left(x_u^2+y_u^2+h_u^2\right)},
    \end{aligned}
    }
    \vspace{-1mm}
\end{equation}
\begin{equation}\label{SIR2}
            \vspace{-3.0mm}
{\small
\begin{aligned}
      \hspace{-8mm}\textrm{SIR}_2(x_{_{\textrm{MSI}}},\hat{y}_{_{\textrm{MSI}}})=
      \hspace{-1mm} \frac{\hspace{-.15mm}p_u\hspace{-.1mm} \left(\hspace{-.15mm}\hat{y}^2_{_{\textrm{MSI}}}\hspace{-.15mm}+\hspace{-.0mm}(\hspace{-.13mm}D-x_{_{\textrm{MSI}}}\hspace{-.13mm})^2\hspace{-.13mm}\right)}{\hspace{-0.1mm}p_{_{\textrm{MSI}}}\hspace{-0.15mm}\left(\left(\hspace{-0.14mm}D\hspace{-0.12mm}-\hspace{-0.12mm}x_u\hspace{-0.13mm}\right)\hspace{-.12mm}^2\hspace{-0.12mm}+y^2_u+\hspace{-0.2mm}h^2_u\hspace{-.2mm}\right)\hspace{-0.0mm} \left(\frac{\hspace{-0.15mm}\eta_{_{\textrm{NLoS}}}}{\hspace{-0.15mm}\mu_{_{\textrm{NLoS}}}}\hspace{-.17mm}\right)},
      \end{aligned}
      }
\end{equation}

\begin{equation}\label{SIR3}
{\small
\begin{aligned}
      \hspace{1.8mm}\textrm{SIR}_3(x_{_{\textrm{MSI}}},\hat{y}_{_{\textrm{MSI}}})=
      \hspace{-1mm}\frac{p_{_{\textrm{TR\_2}}} \hspace{-.12mm} \left(\hspace{-.15mm}\left(\hspace{-.15mm}x_u-x_{_{\textrm{MSI}}}\hspace{-.15mm}\right)^2\hspace{-.15mm}+\hspace{-.14mm}\left(\hat{y}_{_{\textrm{MSI}}}-y_u\right)^2\hspace{-.15mm}+\hspace{-.12mm}h_u^2\hspace{-.15mm}\right)}{p_{_{\textrm{MSI}}}\left((D-x_u)^2+y^2_u+h_u^2\right)},
      \end{aligned}
      }
\end{equation}
\begin{equation}\label{SIR4}
{\small
\begin{aligned}
     \hspace{-16mm} \textrm{SIR}_4(x_{_{\textrm{MSI}}},\hat{y}_{_{\textrm{MSI}}})=
      \hspace{-1mm} \frac{\hspace{-.15mm}p_u\hspace{-.1mm} \left(\hspace{-.15mm}\hat{y}^2_{_{\textrm{MSI}}}\hspace{-.15mm}+\hspace{-.0mm}x_{_{\textrm{MSI}}}^2\hspace{-.13mm}\right)}{\hspace{-0.1mm}p_{_{\textrm{MSI}}}\hspace{-0.15mm}\left(x_u^2\hspace{-0.12mm}+y^2_u+\hspace{-0.2mm}h^2_u\hspace{-.2mm}\right)\hspace{-0.0mm} \left(\frac{\hspace{-0.15mm}\eta_{_{\textrm{NLoS}}}}{\hspace{-0.15mm}\mu_{_{\textrm{NLoS}}}}\hspace{-.17mm}\right)}.
      \end{aligned}
      }
            \vspace{-2.0mm}
\end{equation}
Consequently, the SIR of Link\_1 and Link\_2 are given by:
\vspace{-2mm}
\begin{equation}\label{SIR_L_1}
\hspace{-11mm}{\small
\begin{aligned}
 \textrm{SIR}_{\textrm{Link\_1}}(x_{_{\textrm{MSI}}},\hat{y}_{_{\textrm{MSI}}})=\min\{&\textrm{SIR}_{1}(x_{_{\textrm{MSI}}},\hat{y}_{_{\textrm{MSI}}}),\textrm{SIR}_{2}(x_{_{\textrm{MSI}}},\hat{y}_{_{\textrm{MSI}}}) \},
 \end{aligned}
 }\hspace{-10mm}
     \vspace{-1mm}
\end{equation}
\begin{equation}\label{SIR_L_2}
\hspace{-12mm}{\small
\begin{aligned}
    \textrm{SIR}_{\textrm{Link\_2}}(x_{_{\textrm{MSI}}},\hat{y}_{_{\textrm{MSI}}})=\min\{&\textrm{SIR}_{3}(x_{_{\textrm{MSI}}},\hat{y}_{_{\textrm{MSI}}}),\textrm{SIR}_{4}(x_{_{\textrm{MSI}}},\hat{y}_{_{\textrm{MSI}}}) \},
    \end{aligned}
    }\hspace{-10mm}
    \vspace{-1mm}
\end{equation}
and the optimal position of the jammer is given by:
{\small
\begin{equation}\label{MainProb}
\hspace{-22mm}
    x^*_{_{\textrm{MSI}}}\hspace{-1mm}=\hspace{-5mm}\argmin_{-x^-_{jam}\leq x_{_{\textrm{MSI}}}\leq x^+_{jam} }\hspace{-3mm} \max \{\hspace{-.4mm}\textrm{SIR}_{\textrm{Link\_1}}(\hspace{-.4mm}x_{_{\textrm{MSI}}},\hat{y}_{_{\textrm{MSI}}}\hspace{-.4mm}),\hspace{-.4mm}\textrm{SIR}_{\textrm{Link\_2}}(\hspace{-.4mm}x_{_{\textrm{MSI}}},\hat{y}_{_{\textrm{MSI}}}\hspace{-.4mm}) \hspace{-.4mm}\}.
    \hspace{-18mm}
        \vspace{-1mm}
\end{equation}
}
  As discussed earlier, $\textrm{SIR}_{\textrm{Link\_1}}$ and $\textrm{SIR}_{\textrm{Link\_2}}$ are, in general, non-convex functions w.r.t $x_{_{\textrm{MSI}}}$. This results in non-convexity of~\eqref{MainProb}. The direct approach to solve~\eqref{MainProb} is to obtain the  mathematical expressions of $\textrm{SIR}_{\textrm{Link\_1}}$ and $\textrm{SIR}_{\textrm{Link\_2}}$, and then solve~\eqref{MainProb} using a non-convex optimization technique. However, functions $\textrm{SIR}_{\textrm{Link\_1}}$ and $\textrm{SIR}_{\textrm{Link\_2}}$ are \textit{piecewise-defined functions}.\footnote{A piecewise-defined function is a function defined by multiple sub-functions, each of which applying to a certain interval of the original function's domain.} This makes  $\textrm{SIR}_{\textrm{max}}$ a piece-wise function, for which the detailed specification is tedious. Also, it can be noticed that upon having multiple UAVs this approach is intractable. Considering this fact, we propose a systematic approach to efficeintly obtain the solution of~\eqref{MainProb}.

\begin{definition}
A function $f:\mathbb{R}\rightarrow\mathbb{R}$ is called a \textbf{piecewise convex} function if it can be represented as $f(x)=\min\{f_j(x): j\in \mathcal{M}\}$, where $f_j:\mathbb{R}\rightarrow\mathbb{R}$ is convex $\forall j \in \mathcal{M}\triangleq \{1,2,\cdots,|\mathcal{M}|\}$.
\end{definition}
In other words, the domain of a piecewise convex function can be partitioned into multiple intervals such that at each interval the corresponding sub-function is convex. Note that piecewise convex functions are in general non-convex.
In the following, we present three lemmas, the proofs of which are straightforward and omitted in the interest of space. All of the functions considered below are assumed to be continuous.

\begin{lemma}\label{lemmaCrit1}
Let $g_i:\mathbb{R}\rightarrow\mathbb{R}$, $1\leq i\leq M$, be convex functions with the set of \textit{critical points} $\mathcal{C}_{g_i}$, $\forall i$.\footnote{At any critical point such as $(x,g(x))$, the derivative of the function $g$ is either zero or does not exist.} Function $q=\min(g_1,\cdots,g_M)$ is a piecewise convex function, for which the set of critical points $\mathcal{C}_q$ is given by: $\mathcal{C}_q\subset \displaystyle \left( \underset{i:1\leq i\leq M}{\cup} \mathcal{C}_{{g_i}}\right)\cup \left( \underset{{(i,j): 1\leq i <j \leq M}}{\cup} \mathcal{S}_{g_i,g_j}\right)$, where $\mathcal{S}_{g_i,g_j}\triangleq \{(x,g_i(x)): x\in\mathbb{R},~g_i(x)=g_j(x)  \}$. 
\end{lemma}

\begin{lemma}\label{lemmaCrit2}
Let $z_i:\mathbb{R}\rightarrow\mathbb{R}$, $1\leq i\leq M$, be piecewise convex functions with the set of critical points $\mathcal{C}_{z_i}$, $\forall i$. Function $w=\max (z_1,\cdots,z_M)$ is a piecewise convex function, for which the set of critical points $\mathcal{C}_w$ is given by: $\mathcal{C}_w\subset \displaystyle \left( \underset{i:1\leq i\leq M}{\cup} \mathcal{C}_{{z_i}}\right) \cup \left( \underset{{(i,j): 1\leq i <j \leq M}}{\cup} \mathcal{S}_{z_i,z_j}\right)$, where $\mathcal{S}_{z_i,z_j}\triangleq \{(x,z_i(x)): x\in\mathbb{R},~z_i(x)=z_j(x)  \}$.
\end{lemma}

\begin{lemma}\label{lemmaCrit3}\hspace{-1.3mm}Let \hspace{-0.5mm}$f\hspace{-.3mm}:\hspace{-.1mm}\mathbb{R}\hspace{-.2mm}\rightarrow\hspace{-.2mm}\mathbb{R}$\hspace{-0.3mm} be a piecewise convex function with the set of critical points $\mathcal{C}_f$. The global minimum of the function $\left(x^*_f, f(x^*_f)\right)$, where $x^*_f=\argmin_{x\in\mathbb{R}}f(x)$, always belongs to the set of critical points of the function, i.e., $\left(x^*_f, f(x^*_f)\right)\in \mathcal{C}_f$.
\end{lemma}

In other words, in a special case where $M=2$, Lemma~\ref{lemmaCrit1} asserts that the critical points of function $q=\min(g_1,g_2)$, where $g_1$ and $g_2$ are two convex functions, are either located at the intersections of $g_1$ and $g_2$ or coincide with those of $g_1$ and $g_2$. Lemma~\ref{lemmaCrit2} conveys a similar message for the maximum of two piecewise convex functions. Also, according to Lemma~\ref{lemmaCrit3} the minimum of a piecewise convex function is always among the critical points of the function.  Given the convexity (and continuity) of~\eqref{SIR1}-\eqref{SIR4}, $\textrm{SIR}_{\textrm{Link\_1}}$ and $\textrm{SIR}_{\textrm{Link\_2}}$ are both piecewise continuous convex functions w.r.t $x_{_{\textrm{MSI}}}$, which results in the piecewise convexity of $\textrm{SIR}_{\textrm{max}}$. According to Lemma~\ref{lemmaCrit3}, the global minimum of $\textrm{SIR}_{\textrm{max}}$, i.e., the solution of~\eqref{MainProb}, belongs to its set of critical points $\mathcal{C}_{\textrm{SIR}_{\textrm{max}}}$, which is a subset of the set of critical points of $\textrm{SIR}_{\textrm{Link\_1}}$ and $\textrm{SIR}_{\textrm{Link\_2}}$, i.e., $\mathcal{C}_{\textrm{SIR}_{\textrm{Link\_1}}}$ and $\mathcal{C}_{\textrm{SIR}_{\textrm{Link\_2}}}$, and the intersection points of $\textrm{SIR}_{\textrm{Link\_1}}$ and $\textrm{SIR}_{\textrm{Link\_2}}$. However, direct derivation of the intersection points of $\textrm{SIR}_{\textrm{Link\_1}}$ and $\textrm{SIR}_{\textrm{Link\_2}}$ requires obtaining their expressions, which we aim to avoid. In the following, we present a corollary that alleviates this issue.

\begin{corollary}\label{cor:1}\hspace{-1.4mm}
Let $v_1=\min (f_1,\cdots,f_{_{N+1}})$ and $v_2=\min (f_{_{N+1}},\cdots,f_{_{2N+2}})$, where $f_i$, $\forall i$, is a single variable convex function with its domain and range defined on the set of real numbers, and let $v=\max(v_1,v_2)$. Then, \vspace{-2.1mm}\begin{equation}
    \mathcal{C}_v \subset \displaystyle \left(\underset{i:1\leq i\leq 2N+2}{\cup} \mathcal{C}_{{f_i}} \right) \cup \left( \underset{{(i,j): 1\leq i <j \leq 2N+2}}{\cup} \mathcal{S}_{f_i,f_j}\right),
    \vspace{-2mm}
\end{equation} where $\mathcal{S}_{f_i,f_j}\triangleq \{(x,f_i(x)): x\in\mathbb{R},~f_i(x)=f_j(x)  \}$. Let $\Psi_v=\left(\underset{i:1\leq i\leq 2N+2}{\cup} \mathcal{C}_{{f_i}}\right) \cup \left( \underset{{(i,j): 1\leq i <j \leq 2N+2}}{\cup} \mathcal{S}_{f_i,f_j}\right)$ denote the \textit{candidate set of critical points} of function $v$. The global minimum of the piecewise convex function $v$, i.e., $(x^*_v,v(x^*_v))$, where $x^*_v=\argmin_{x\in \mathbb{R}} v(x)$, can be found as follows:
\vspace{-1.5mm}
\begin{equation}\label{eq:minFinder}
    x^*_v= \argmin_{x} \{v(x): (x,y)\in\Psi_v, v(x)=y \}.
       \vspace{-1.5mm}
\end{equation}
\end{corollary}
In Corollary~\ref{cor:1}, we reveal a fast method of obtaining the minimum of the piecewise function $v$ as defined above, by solely inspecting the points belonging to the candidate set of critical points. In the following, we first derive the candidate set of critical points of function $\textrm{SIR}_{\textrm{max}}$ and then propose an algorithm that implements Corollary~\ref{cor:1} assuming $N=1$ with $f_i=\textrm{SIR}_i(x_{_{\textrm{MSI}}},\hat{y}_{_{\textrm{MSI}}})$, $i\in\{1,2,3,4\}$,  $v_1=\textrm{SIR}_{\textrm{Link\_1}}$ and $v_2=\textrm{SIR}_{\textrm{Link\_2}}$ to obtain the minimum of $\textrm{SIR}_{\textrm{max}}$.

\begin{proposition}\label{prop:exterma} The critical point of function $\textrm{SIR}_1(x_{_{\textrm{MSI}}},\hat{y}_{_{\textrm{MSI}}})$, $\textrm{SIR}_2(x_{_{\textrm{MSI}}},\hat{y}_{_{\textrm{MSI}}})$, $\textrm{SIR}_3(x_{_{\textrm{MSI}}},\hat{y}_{_{\textrm{MSI}}})$, $\textrm{SIR}_4(x_{_{\textrm{MSI}}},\hat{y}_{_{\textrm{MSI}}})$ is
 $x^{(1)}_{_{\textrm{MSI}}} = x_u$, $x^{(2)}_{_{\textrm{MSI}}} = D$, $x^{(3)}_{_{\textrm{MSI}}} = x_u$, and $x^{(4)}_{_{\textrm{MSI}}} = 0$, respectively.
\end{proposition}
\begin{lemma}\label{lem:LemaaMother}
Consider the equality of two distinct quadratic curves in  the format of $A\left[(x-B)^2+C\right]=D\left[(x-E)^2+F\right]$. Define $\Delta \triangleq (AB-DE)^2+(D-A)\left[A(B^2+C)-D(F+E^2)\right]$. If $\Delta<0$, the quadratic equations have no intersection; otherwise, the intersecting points are given by:\footnote{Throughout, we use sub-index $+$ and $-$ to denote the larger and the smaller solution, respectively. Note that if $\Delta=0$, $x_-=x_+$.}
\vspace{-1mm}
{\small
\begin{equation}\label{LemaaMother}
\begin{aligned}
   & x_{\pm}=\frac{AB-DE\pm \sqrt{\Delta}}{A-D}~~\textrm{if}~~ A\neq D,\\
   & x_-=x_+=\frac{(B^2+C)-(E^2+F)}{2(B-E)}~~\textrm{O.W.}
    \end{aligned}
    \vspace{-2mm}
\end{equation}
}
\end{lemma}
\begin{proposition}\label{prop:crosspointsSinlge}
The intersection points of the two SIR curves as a function of $x_{_{\textrm{MSI}}}$ for each link in the dual-hop setting are given as follows, where $x^{(i,j)}_{\pm}$ denote the intersection points of $\textrm{SIR}_i$ and $\textrm{SIR}_j$:



\begin{itemize}
\setlength{\itemindent}{-1em}
\item 
For Link\_1, replace {\small $A=p_{_{\textrm{TR\_1}}}/\Big[x^2_u+y^2_u+h^2_u\Big]$, $B=x_u$, $C=(\hat{y}_{_\textrm{MSI}}-y_u)^2+h_u^2$, $D=p_u/\Big[((D-x_u)^2+y^2_u+h^2_u)\left(\frac{\eta_{_{\textrm{NLoS}}}}{\mu_{_{\textrm{NLoS}}}}\right)\Big]$, $E=D$,} and {\small$F=\hat{y}^2_{_\textrm{MSI}}$} in \eqref{LemaaMother} to obtain {\small$x^{(1,2)}_{\pm}$}.

\item 
For Link\_2, replace {\small$A=p_{_{\textrm{TR\_2}}}/\Big[{((D-x_u)^2+y^2_u+h^2_u)}\Big]$, $B=x_u$, $C=(\hat{y}_{_\textrm{MSI}}-y_u)^2+h_u^2$, $D=p_u/\Big[(x^2_u+y^2_u+h^2_u)\left(\frac{\eta_{_{\textrm{NLoS}}}}{\mu_{_{\textrm{NLoS}}}}\right)\Big]$, $E=0$,} and {\small$F=\hat{y}^2_{_\textrm{MSI}}$} in \eqref{LemaaMother} to obtain {\small$x^{(3,4)}_{\pm}$}.
\end{itemize}
\end{proposition}
\begin{proposition}\label{prop:crosspointsDouble}
 The four SIR curves in the dual-hop setting intersect with each other in the following points:
 \vspace{-1mm}
\begin{itemize}
\setlength{\itemindent}{-1em}
    \item If $\frac{p_{_{\textrm{TR\_1}}} ((D-x_x)^2+y^2_u+h^2_u)}{p_{_{\textrm{TR\_2}}}(x_u^2+y^2_u+h^2_u)}=1$, two functions $\textrm{SIR$_1$}$, $\textrm{SIR$_3$}$ are always equal; otherwise, they have no intersection.\footnote{\vspace{-.1mm}When  the two functions match, their critical points also match. Hence, we can easily assume that $x^{(1,3)}_{\pm}$ do not exist without affecting the analysis.} 

\item 
For $\textrm{SIR$_1$}$, $\textrm{SIR$_{4}$}$, replace {\small$A=\frac{p_{_{\textrm{TR\_1}}}}{x^2_u+y^2_u+h^2_u}$, $B=x_u$, $C=(\hat{y}_{_\textrm{MSI}}-y_u)^2+h_u^2$,
$D=p_u/\Big[(x^2_u+y^2_u+h^2_u)\left(\frac{\eta_{_{\textrm{NLoS}}}}{\mu_{_{\textrm{NLoS}}}}\right)\Big]$, $E=0$,} and {\small$F=\hat{y}^2_{_\textrm{MSI}}$} in \eqref{LemaaMother} to obtain {\small$x^{(1,4)}_{\pm}$}.

\item 
For $\textrm{SIR$_2$}$, $\textrm{SIR$_3$}$, replace
{\small$A=p_u/\Big[((D-x_u)^2+y^2_u+h^2_u)\left(\frac{\eta_{_{\textrm{NLoS}}}}{\mu_{_{\textrm{NLoS}}}}\right)\Big]$, $B=D$, $C=\hat{y}^2_{_\textrm{MSI}}$,
$D=\frac{p_{_{\textrm{TR\_2}}}}{((D-x_u)^2+y^2_u+h^2_u)}$, $E=x_{_{\textrm{MSI}}}$,} and {\small$F=(\hat{y}_{_\textrm{MSI}}-y_u)^2+h_u^2$} in \eqref{LemaaMother} to obtain $x^{(2,3)}_{\pm}$. 

\item 
For $\textrm{SIR$_2$}$,~$\textrm{SIR$_4$}$, replace $A=p_u/\Big[((D-x_u)^2+y^2_u+h^2_u)\left(\frac{\eta_{_{\textrm{NLoS}}}}{\mu_{_{\textrm{NLoS}}}}\right)\Big]$, $B=D$, $C=\hat{y}^2_{_\textrm{MSI}}$, $D=p_u/\Big[(x^2_u+y^2_u+h^2_u)\left(\frac{\eta_{_{\textrm{NLoS}}}}{\mu_{_{\textrm{NLoS}}}}\right)$, $E=0$, and $F=\hat{y}^2_{_\textrm{MSI}}$ in \eqref{LemaaMother} to obtain $x^{(2,4)}_{\pm}$.

\end{itemize}
\end{proposition}

The pseudo-code of our optimal jammer placement algorithm is given in Algorithm~\ref{alg:jammDual}. The input $\hat{y}_{_{\textrm{MSI}}}$ is inherently assumed, and thus eliminated from the argument of the SIR functions  for compactness. The algorithm uses the candidate set of critical points of function $\textrm{SIR}_{\textrm{max}}$, which consists of the points obtained in Proposition~\ref{prop:exterma},~\ref{prop:crosspointsSinlge}, and~\ref{prop:crosspointsDouble}. Note that in cases where $x_{\pm}^{(i,j)}$ does not exist according to Lemma~\ref{lem:LemaaMother}, the algorithm automatically skips it. For each of the points, the algorithm first tests the feasibility of the point, i.e., $v(x)=y$ in~\eqref{eq:minFinder}. For instance, for $\left(x^{(1)}_{_{\textrm{MSI}}},\textrm{SIR}_1(x^{(1)}_{_{\textrm{MSI}}})\right)$, it checks that this point also belongs to $\textrm{SIR}_{\textrm{max}}$ in lines~\ref{alg:tester} and~\ref{alg:tester2}.
Finally, it derives the minimum of function $\textrm{SIR}_{\textrm{max}}$, i.e., the solution of~\eqref{MainProb}, according to Corollary~\ref{cor:1} by testing all the feasible candidates for the critical points of the function in line~\ref{alg:finalline}. Note that our method  reduces the analysis of an intractable function to systematic calculation of values of the SIR expressions at $14$ points (c.f. Footnote 6).
\vspace{-2mm}
 \section{Jammer Placement in Multi-hop Setting}\label{sec:multiple}
\noindent Consider the system model explained in Section~\ref{sec:bigPicture} and depicted in Fig.~\ref{fig:multiple}. Let $\delta_{x(i,j)}=x_i-x_j$,  $\delta_{y(i,j)}=y_i-y_j$, $\delta_{h(i,j)}=h_i-h_j$ , for $\textrm{UAV}_i$ and $\textrm{UAV}_j$. 
In this case, the SIR expressions for Link\_1 and Link\_2 are given as follows:
\vspace{-2mm}
	    \begin{algorithm}[h]
 	\caption{Optimal jammer placement in dual-hop UAV-assisted relay networks}\label{alg:jammDual}
 	\SetKwFunction{Union}{Union}\SetKwFunction{FindCompress}{FindCompress}
 	\SetKwInOut{Input}{input}\SetKwInOut{Output}{output}
 	 	{\footnotesize
 	The set of final candidates of exterma $\mathcal{P}=\{\}$\\
 	 Derive $x^{(1)}_{_{\textrm{MSI}}}$, $x^{(2)}_{_{\textrm{MSI}}}$, $x^{(3)}_{_{\textrm{MSI}}}$, and $x^{(4)}_{_{\textrm{MSI}}}$ using Proposition~\ref{prop:exterma}.\\
 	 \For{$i\in\{1,2\}$}{
 \If{$\min\{SIR_1(x^{(i)}), SIR_2 (x^{(i)})\} = SIR_i(x^{(i)})$}{
 		\label{alg:tester}  \If{$SIR_i(x^{(i)}) \geq \min\{SIR_3(x^{(i)}), SIR_4 (x^{(i)})\}$}{
 	 \label{alg:tester2} $\mathcal{P}= \mathcal{P} \cup \Big\{\left[x^{(i)}, SIR_i(x^{(i)})\right]\Big\} $}
 		 }
 	 }
 	 \For{$i\in\{3,4\}$}{
 	  	 \If{$\min\{SIR_3(x^{(i)}), SIR_4 (x^{(i)})\} = SIR_i(x^{(i)})$}{
 	 \If{$SIR_i(x^{(i)}) \geq \min\{SIR_1(x^{(i)}), SIR_2 (x^{(i)})\}$}{
 	  $\mathcal{P}= \mathcal{P} \cup \Big\{\left[x^{(i)}, SIR_i(x^{(i)})\right]\Big\} $}
 	 }
 	 }
 	 Derive $x^{(1,2)}_{\pm}$ and $x^{(3,4)}_{\pm}$ using Proposition~\ref{prop:crosspointsSinlge}.\\
 	 Define $y^{(1)}=x^{(1,2)}_{-}$, $y^{(2)}=x^{(1,2)}_{+}$, $z^{(1)}=x^{(3,4)}_{-}$, $z^{(2)}=x^{(3,4)}_{+}$.\\
 	 \For{$i\in\{1,2\}$}{
 	 \If{$SIR_1(y^{(i)}) \geq \min\{SIR_3(y^{(i)}), SIR_4 (y^{(i)})\}$}{
 	  $\mathcal{P}= \mathcal{P} \cup \Big\{\left[y^{(i)}, SIR_1(y^{(i)})\right]\Big\} $}
 	 }
 	 	 \For{$i\in\{1,2\}$}{
 	 \If{$SIR_3(z^{(i)}) \geq \min\{SIR_1(z^{(i)}), SIR_2 (z^{(i)})\}$}{
 	  $\mathcal{P}= \mathcal{P} \cup \Big\{\left[z^{(i)}, SIR_3(z^{(i)})\right]\Big\} $}
 	 }
 	 Derive $x^{(1,4)}_{\pm}$, $x^{(2,3)}_{\pm}$, and $x^{(2,4)}_{\pm}$ using Proposition~\ref{prop:crosspointsDouble}.\\
 	 \For{$(i,j)\in \{ (1,4),(2,3),(2,4)\}$}{
 	 \If{$\min\{SIR_1(x^{(i,j)}_{-}), SIR_2 (x^{(i,j)}_{-})\} = SIR_i(x^{(i,j)}_{-})$ \textbf{and} $\min\{SIR_3(x^{(i,j)}_{-}), SIR_4 (x^{(i,j)}_{-})\} = SIR_j(x^{(i,j)}_{-})$}{
 	  $\mathcal{P}= \mathcal{P} \cup \Big\{\left[ x^{(i,j)}_{-},SIR_i(x^{(i,j)}_{-})\right]\Big\}  $
 	 }
 	 \If{$\min\{SIR_1(x^{(i,j)}_{+}), SIR_2 (x^{(i,j)}_{+})\} = SIR_i(x^{(i,j)}_{+})$ \textbf{and} $\min\{SIR_3(x^{(i,j)}_{+}), SIR_4 (x^{(i,j)}_{+})\} = SIR_j(x^{(i,j)}_{+})$}{
 	 
 	 	  $\mathcal{P}= \mathcal{P} \cup \Big\{\left[ x^{(i,j)}_{+},SIR_i(x^{(i,j)}_{+})\right]\Big\}  $
 	 }}
 	 
 	 Consider $\mathcal{P}$ in the following format: $\mathcal{P}= \cup_{i=1}^{|\mathcal{P}|} \{\left[a_i,b_i \right]\}$\\
 	 $x^*_{\textrm{MSI}}=a_{i^*}$,  $i^*=\displaystyle\argmin_{i} \{b_i: \left[a_i,b_i \right]\in \mathcal{P},  -x^-_{jam}\leq a_i\leq x^+_{jam}  \}$\label{alg:finalline}
 	  }
 	    	  \vspace{-1.0mm}
  \end{algorithm}
  \begin{table*}[b]
  \vspace{-5mm}
\begin{minipage}{0.99\textwidth}
\hrulefill
{\footnotesize{
\begin{equation}\label{longSIR1}
\Phi_{\textrm{SIR}_1}:\left[A=p_{_{\textrm{TR\_1}}}/\Big[x^2_1+y^2_1+h^2_1\Big], B=x_1, C=(\hat{y}_{_\textrm{MSI}}-y_1)^2+h_1^2\right]
\vspace{-2mm}
\end{equation}
\begin{equation}\label{longSIRk}
\Phi_{\textrm{SIR}_{k}}:\left[ A=p_{_{k-1}}\eta_{_{\textrm{NLoS}}}  /\Big[\mu_{_{\textrm{LoS}}}\hspace{-0.5mm}\left( |\delta_ {x_{(k-1,k)}}|^2+|\delta_{y_{(k-1,k)}}|^2+|\delta_{h_{(k-1,k)}}|^2\right)\Big],
    B=x_k, C=(\hat{y}_{_\textrm{MSI}}-y_k)^2+h_k^2\right]~~\textrm{if}~~2\leq k\leq N
    \vspace{-2mm}
\end{equation}
\begin{equation}\label{longSIRNplus1}
\Phi_{\textrm{SIR}_{N+1}}:\left[ A=p_{_N}\mu_{_{\textrm{NLoS}}} /\left(\eta_{_{\textrm{NLoS}}}\left((x_{N}-D)^2+y^2_{N}+h_{N}^2\right)\right), B=D, C=\hat{y}^2_{_\textrm{MSI}}\right]
\vspace{-2mm}
\end{equation}
\begin{equation}\label{longSIRNplus2}
\Phi_{\textrm{SIR}_{N+2}}:\left[ A= p_{_{\textrm{TR\_2}}}/\Big[\left((x_{N}-D)^2+y^2_{N}+h_{N}^2\right)\Big],
    B=x_N,
    C=(\hat{y}_{_\textrm{MSI}}-y_N)^2+h_N^2\right]
    \vspace{-2mm}
\end{equation}
\begin{equation}\label{longSIRNpluskplus2}
\Phi_{\textrm{SIR}_{N+k+2}}: \Big[ A=\frac{p_{_{N-k+1}}\eta_{_{\textrm{NLoS}}}} {\mu_{_{\textrm{LoS}}}\hspace{-0.5mm}\left( |\delta_ {x_{(N-k,N-k+1)}}|^2+|\delta_{y_{(N-k,N-k+1)}}|^2+|\delta_{h_{(N-k,N-k+1)}}|^2\right)},
    B=x_{N-k}, 
    C=(\hat{y}_{_\textrm{MSI}}-y_{N-k})^2+h_{N-k}^2\Big]~~\textrm{if}~~1\leq k\leq N-1
    \vspace{-2mm}
\end{equation}
\begin{equation}\label{longSIR2Nplus2}
\Phi_{\textrm{SIR}_{2N+2}}: \left[A=  p_{_1}\mu_{_{\textrm{NLoS}}}/\Big[{\eta_{_{\textrm{NLoS}}}\left(x_1^2+y^2_1+h_1^2\right)}\Big],B=0, C=\hat{y}^2_{_\textrm{MSI}}\right]
\vspace{-2mm}
\end{equation}
}}
\hrulefill
\end{minipage}
\end{table*}
\begin{equation}\label{eq:SIRsLtR}
\vspace{-8mm}
\hspace{-0.1mm}
\small{
\begin{aligned}
  & \textrm{SIR}_1(x_{_{\textrm{MSI}}},\hat{y}_{_{\textrm{MSI}}})= \frac{p_{_{\textrm{TR\_1}}}\left((x_{_{\textrm{MSI}}}-x_1)^2+(\hat{y}_{_\textrm{MSI}}-y_1)^2+h_1^2\right)}{p_{_{\textrm{MSI}}}\left(x_1^2+y^2_1+h_1^2\right)},\\ 
    \vdots\\\vspace{-1mm}
  & \textrm{SIR}_N(x_{_{\textrm{MSI}}},\hat{y}_{_{\textrm{MSI}}})=\hspace{-0.5mm}\frac{p_{_{N-1}}\eta_{_{\textrm{NLoS}}} \left((x_{_{\textrm{MSI}}}-x_{_N})^2+(\hat{y}_{_\textrm{MSI}}-y_{_N})^2+h_{_N}^2\right) }{\hspace{-3.1mm}p_{_{\textrm{MSI}}}\mu_{_{\textrm{LoS}}}\hspace{-0.5mm}\left( |\delta_ {x_{(N-1,N)}}|^2\hspace{-0.5mm}+\hspace{-0.5mm}|\delta_{y_{(N-1,N)}}|^2\hspace{-0.5mm}+\hspace{-0.5mm}|\delta_{h_{(N-1,N)}}|^2\hspace{-0.5mm}\right)},\\ \vspace{-1mm}
  & \textrm{SIR}_{N+1}(x_{_{\textrm{MSI}}},\hat{y}_{_{\textrm{MSI}}})=\frac{p_{_N}\mu_{_{\textrm{NLoS}}}\left((x_{_{\textrm{MSI}}}-D)^2+\hat{y}^2_{_\textrm{MSI}}\right) }{p_{_{\textrm{MSI}}}\eta_{_{\textrm{NLoS}}}\left((x_{_N}-D)^2+y^2_{_N}+h_{_N}^2\right)},\\
  & \textrm{SIR}_{N+2}(x_{_{\textrm{MSI}}},\hat{y}_{_{\textrm{MSI}}})= \frac{p_{_{\textrm{TR\_2}}}\left((x_{_{\textrm{MSI}}}-x_{_N})^2+(\hat{y}_{_\textrm{MSI}}-y_{_N})^2+h_{_N}^2\right)  }{p_{_{\textrm{MSI}}}\left((x_{_N}-D)^2+y^2_{_N}+h_{_N}^2\right)},\\ 
    \vdots\\\vspace{-1mm}
  & \textrm{SIR}_{2N+1}(x_{_{\textrm{MSI}}},\hat{y}_{_{\textrm{MSI}}})=\hspace{-0.5mm}\frac{p_{_2}\eta_{_{\textrm{NLoS}}} \left((x_{_{\textrm{MSI}}}-x_1)^2+(\hat{y}_{_\textrm{MSI}}-y_1)^2+h_1^2\right) }{\hspace{-1.5mm}p_{_{\textrm{MSI}}}\mu_{_{\textrm{LoS}}}\hspace{-0.5mm}\left( |\delta_ {x_{(1,2)}}|^2+|\delta_{y_{(1,2)}}|^2+|\delta_{h_{(1,2)}}|^2\right)},\\ \vspace{-1mm}
  & \textrm{SIR}_{2N+2}(\mathbf{d},h)=\frac{p_{_1}\mu_{_{\textrm{NLoS}}}\left(x^2_{_{\textrm{MSI}}}+\hat{y}^2_{_\textrm{MSI}}\right) }{p_{_{\textrm{MSI}}}\eta_{_{\textrm{NLoS}}}\left(x_1^2+y^2_1+h_1^2\right)}.
    \end{aligned}}
 \end{equation}  
 Similar to Section~\ref{sec:singleUAV}, our method is based on Corollary~\ref{cor:1}. In the following, we derive the candidate set of critical points of function $\textrm{SIR}_{\textrm{max}}$.
 \vspace{-1mm}
 \begin{proposition}\label{prop:extermaMulti} Define $x_0=0$ and $x_{N+1}=D$. For Link\_1, the critical points of the functions  $\textrm{SIR}_k(x_{_{\textrm{MSI}}},\hat{y}_{_{\textrm{MSI}}})$, $1\leq k\leq N+1$, are $x^{(k)}_{_{\textrm{MSI}}} = x_i$.
 Also, for Link\_2, the critical points of 
 $\textrm{SIR}_{N+k+2}(x_{_{\textrm{MSI}}},\hat{y}_{_{\textrm{MSI}}})$, $0\leq k\leq N$, are
 $x^{(N+k+2)}_{_{\textrm{MSI}}} = x_{N-k}$.
\end{proposition}

\begin{proposition}\label{prop:crosspointsSinlgeMulti}
Consider the set of coefficients corresponding to  $\Phi_{\textrm{SIR}_{1}}$,$\Phi_{\textrm{SIR}_{k}}$,$\Phi_{\textrm{SIR}_{N+1}}$,$\Phi_{\textrm{SIR}_{N+2}}$,$\Phi_{\textrm{SIR}_{N+k+2}}$, and $\Phi_{\textrm{SIR}_{2N+2}}$ given in \eqref{longSIR1}-\eqref{longSIR2Nplus2}. To obtain the intersections of the SIR curves of Link\_1,  substitute $\Phi_{\textrm{SIR}_{j}}$ and $\Phi_{\textrm{SIR}_{j'}}$, $1\leq j<j'\leq N+1$, in \eqref{LemaaMother} to obtain $x^{(j,j')}_{\pm}$.
For Link\_2, substitute $\Phi_{\textrm{SIR}_{N+j+2}}$ and $\Phi_{\textrm{SIR}_{N+j'+2}}$, $0\leq j<j'\leq N$, in \eqref{LemaaMother} to obtain $x^{(N+j+2,N+j'+2)}_{\pm}$.
\end{proposition}

\begin{proposition}\label{prop:crosspointsDoubleMulti}
Consider the set of coefficients given in \eqref{longSIR1}-\eqref{longSIR2Nplus2}. To obtain the intersections of the SIR curves of Link\_1 and Link\_2, substitute $\Phi_{\textrm{SIR}_{j}}$, $1\leq j\leq N+1$, and $\Phi_{\textrm{SIR}_{N+j'+2}}$, $0\leq j'\leq N$, in \eqref{LemaaMother} to obtain $x^{(j,N+j'+2)}_{\pm}$.
\end{proposition}
\vspace{-1mm}
The pseudo-code of our optimal jammer placement algorithm in the multi-hop relaying setting is given in Algorithm~\ref{alg:jammeMulti}. As before,  the input $\hat{y}_{_{\textrm{MSI}}}$ is inherently assumed and eliminated from the argument of the SIR functions for compactness. The logic and steps of the algorithm are similar to Algorithm~\ref{alg:jammDual}, and thus we avoid further explanations. It is noteworthy to mentioned that, for $N\geq2$ UAVs, using our method, obtaining the position of the jammer is reduced to systematic calculation of values of SIR expressions at $4N^2+8N+4\sim O(N^2)$ points, which is tractable even in large-scale networks.\footnote{This is the sum of the points given by Proposition~\ref{prop:extermaMulti}, which is $2N+2$, Proposition~\ref{prop:crosspointsSinlgeMulti}, which is $2N(N+1)$, and Proposition~\ref{prop:crosspointsDoubleMulti}, which is $2(N+1)^2$. In the dual-hop setting ($N=1$), only $14$ points need to be examined. This due to the reciprocity of the SIR expressions that eliminates two solutions (see the first case of Proposition~\ref{prop:crosspointsDouble}).}
\section{Simulation Results}\label{simres}
\begin{figure*}[h]
  \centering
\minipage{0.23\textwidth}
  \includegraphics[width=\linewidth]{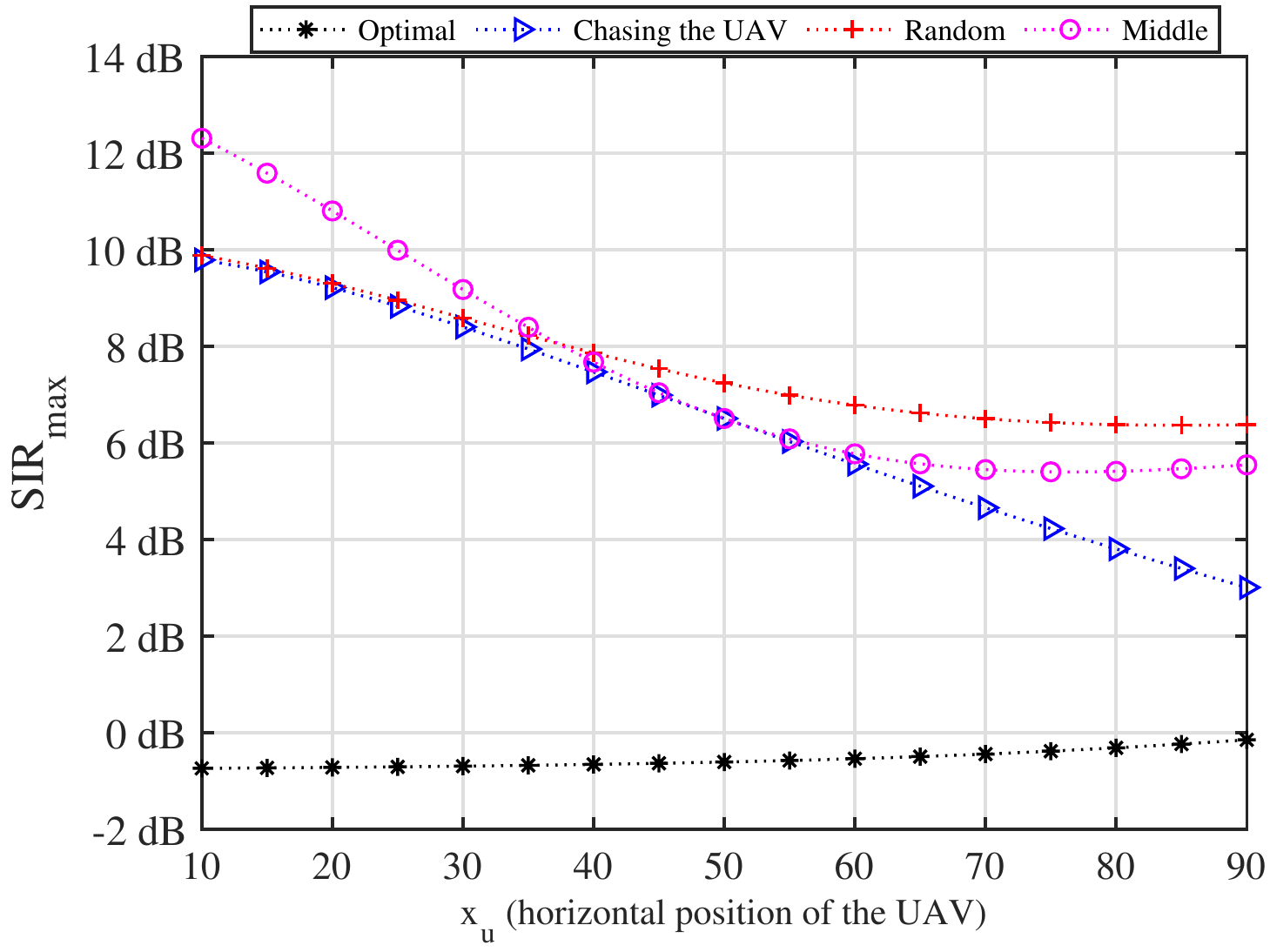}
  \caption{Comparison between SIR$_{\textrm{max}}$ considering moving the UAV in the interval $x_u\in[10,90]$ upon using our optimal method as compared to the baseline methods. }\label{fig:1}
\endminipage\hfill
\minipage{0.23\textwidth}
  \includegraphics[width=\linewidth]{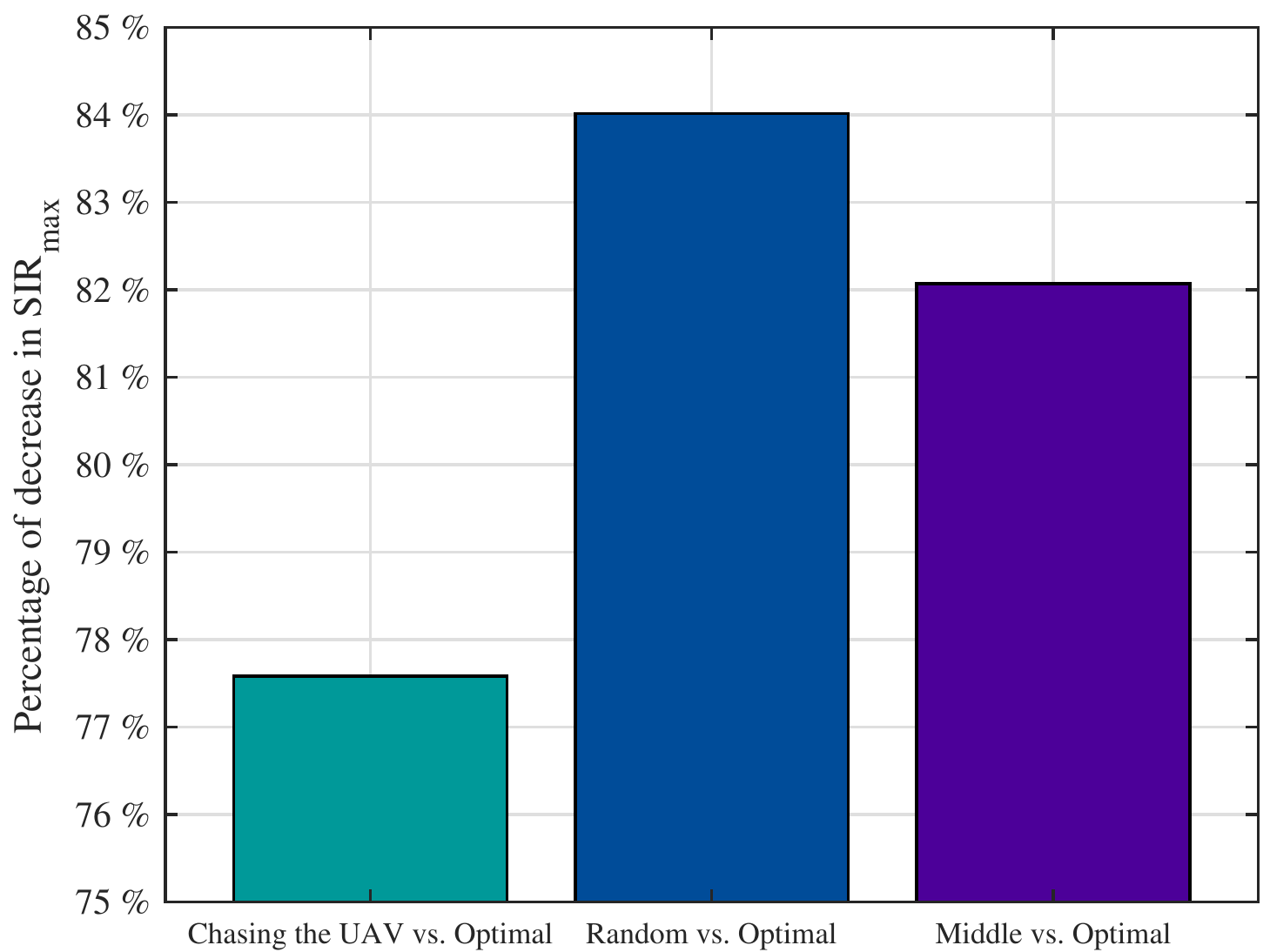}
  \caption{Average percentage of decrease in SIR$_{\textrm{max}}$ considering moving the UAV in the interval $x_u\in[10,90]$ upon using our optimal method as compared to the baseline methods.}\label{fig:2}
\endminipage\hfill
\minipage{0.23\textwidth}%
  \includegraphics[width=\linewidth]{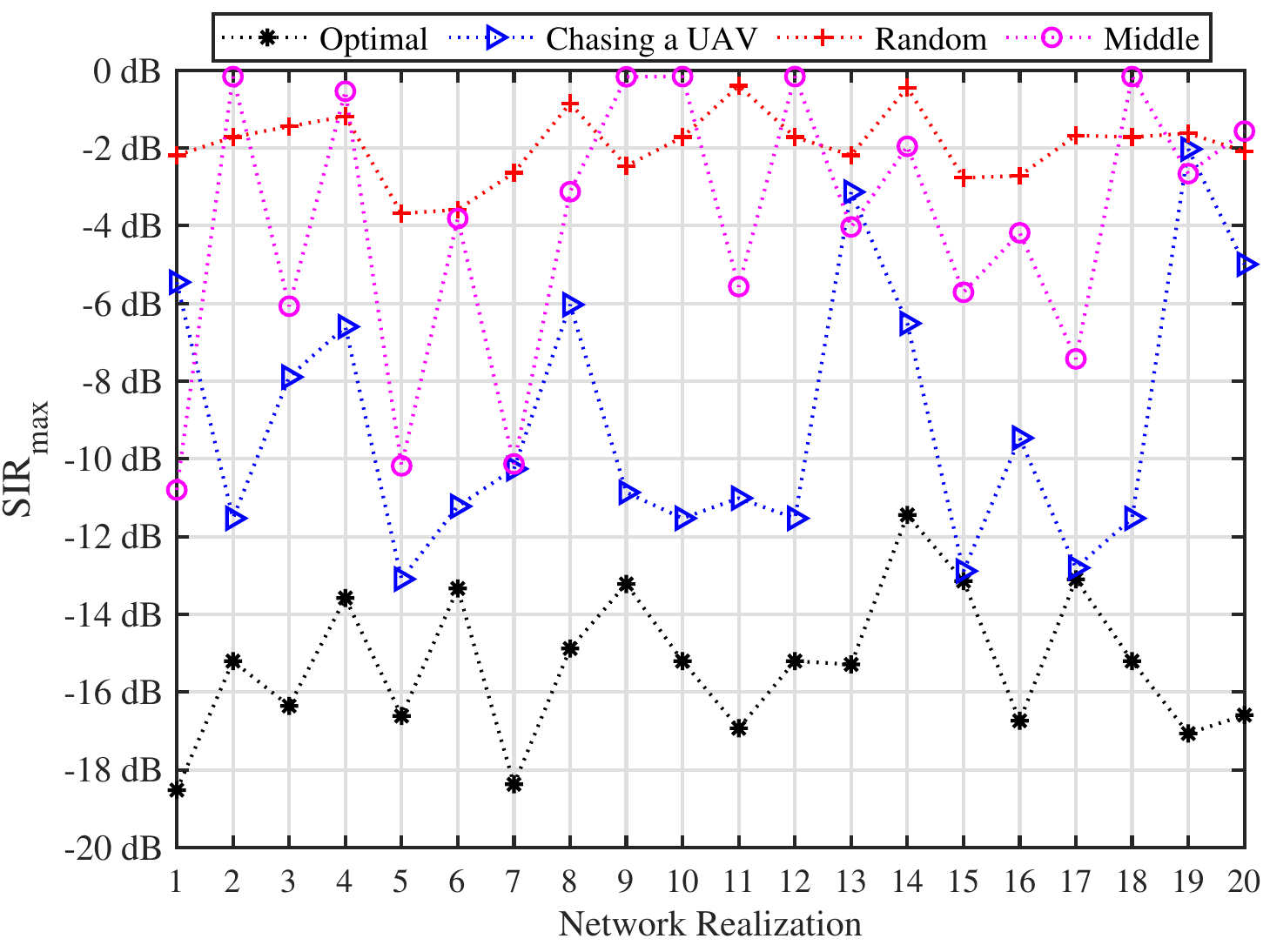}
  \caption{Comparison between SIR$_{\textrm{max}}$ considering $20$ network realizations upon using our optimal method as compared to the baseline methods for $20$ UAV relays in the network.}\label{fig:3}
\endminipage\hfill
\minipage{0.23\textwidth}
  \includegraphics[width=\linewidth]{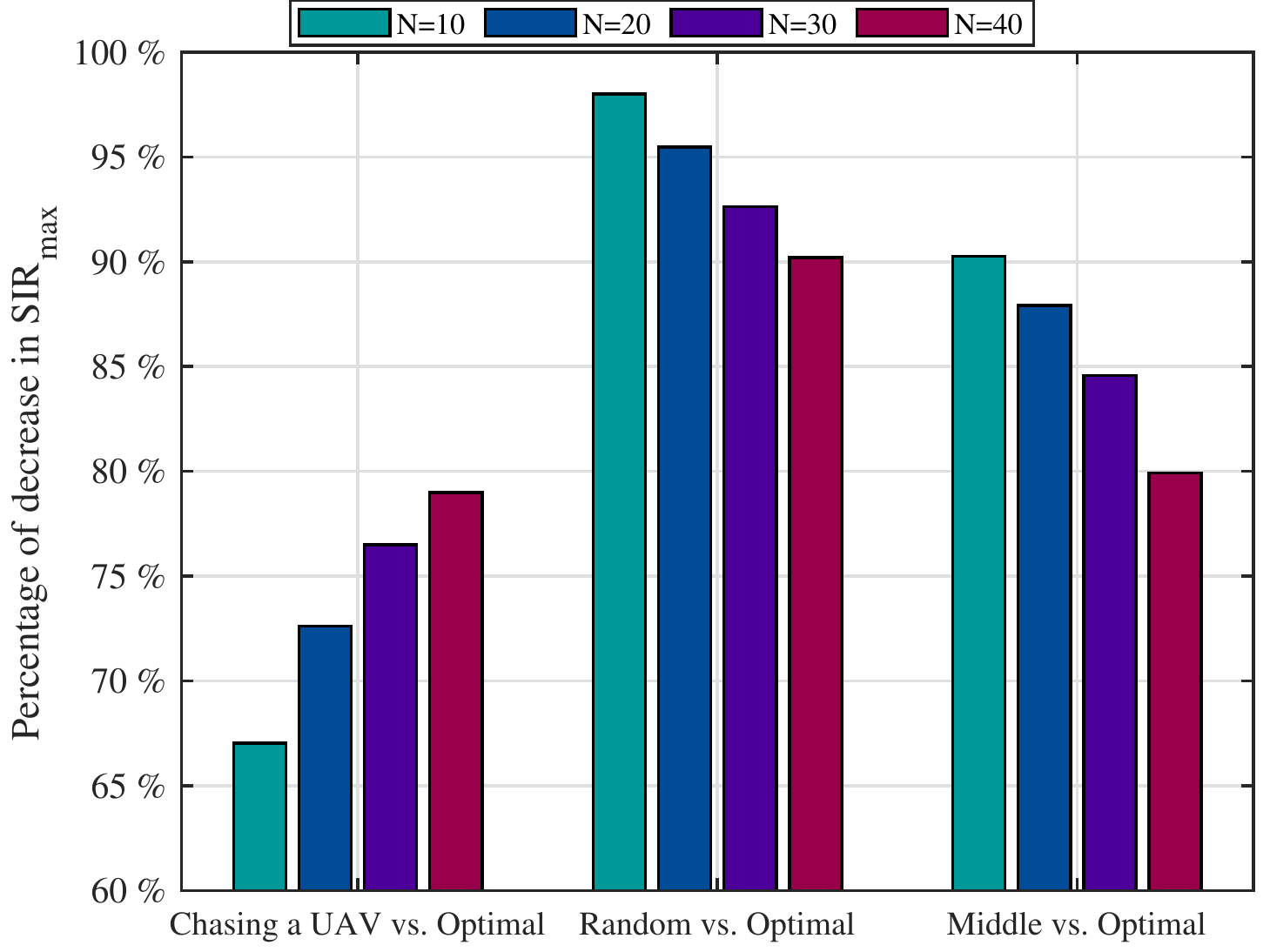}
  \caption{Average percentage of decrease in SIR$_{\textrm{max}}$ considering $300$ network realizations upon using our optimal method as compared to the baseline methods for different numbers of UAV relays in the network~$\small N$.}\label{fig:4}
\endminipage
\vspace{-7mm}
\end{figure*}

\vspace{-1mm}
\noindent Similar to \cite{mozaffari2017mobile}, we consider $f_c=2\textrm{GHz}$, $C_{_{\textrm{LoS}}}=3 \textrm{dB}$, $C_{_{\textrm{NLoS}}}=23\textrm{dB}$, and $\eta_{_{\textrm{NLoS}}}=\mu_{_{\textrm{LoS}}}$. Also, we assume $p_{_{\textrm{MSI}}}=20 \textrm{dBm}$, $p_{_{\textrm{TR\_1}}}=30 \textrm{dBm}$, and $p_{_{\textrm{TR\_2}}}=20 \textrm{dBm}$. Since, considering our network setting, we are among the first to study the jammer placement, we propose the following baselines for performance comparison: i) \textit{Chasing a UAV}: the jammer is placed directly under a UAV relay. ii) \textit{Random}: the jammer is placed in a random position between the TRs. iii) \textit{Middle}: The jammer is placed at the middle of the line between the TRs. Considering the dual-hop setting with $ p_u=20 \textrm{dBm}$, $h_u=45\textrm{m}$, $D=100\textrm{m}$, and $y_u=0\textrm{m}$, Fig.~\ref{fig:1} depicts $\textrm{SIR}_{\textrm{max}}$ upon moving the UAV from $x_u=10\textrm{m}$ to $x_u=90\textrm{m}$.  As can be seen, the best baseline  method is chasing

 \begin{algorithm}[!t]
   \vspace{-.2mm}
 	\caption{Optimal jammer placement in multi-hop UAV-assisted relay networks}\label{alg:jammeMulti}
 	\SetKwFunction{Union}{Union}\SetKwFunction{FindCompress}{FindCompress}
 	\SetKwInOut{Input}{input}\SetKwInOut{Output}{output}
 	 	{\footnotesize
 	The set of final candidates of exterma $\mathcal{P}=\{\}$\\
 	 Derive $x^{(k)}_{_{\textrm{MSI}}}$, $1\leq k\leq 2N+2$ using Proposition~\ref{prop:extermaMulti}.\\
 	 \For{$i\in\{1,2,\cdots,N+1\}$}{
 	 \If{$\min\{SIR_j(x^{(i)}): 1\leq j\leq N+1\} = SIR_i(x^{(i)})$}{
 	 \If{\hspace{-.5mm}$SIR_i(x^{(i)})\hspace{-.5mm} \geq \hspace{-.5mm}\min\{ SIR_{N+1+j}(x^{(i)}):\hspace{-.5mm} 1\hspace{-.5mm}\leq\hspace{-.5mm} j\hspace{-.5mm}\leq\hspace{-.5mm} N+1\}$}{  \vspace{-1mm}
 	 $\mathcal{P}= \mathcal{P} \cup \Big\{\left[x^{(i)}, SIR_i(x^{(i)})\right]\Big\} $
 	 \vspace{-1mm}}
 	 }
 	 }
 	 \For{$i\in\{N+2,N+3,\cdots,2N+2\}$}{
 	  	 \If{$\min\{SIR_{N+1+j}(x^{(i)}): 1\leq j\leq N+1\} = SIR_i(x^{(i)})$}{
 	 \If{$SIR_i(x^{(i)}) \geq \min\{SIR_j(x^{(i)}): 1\leq j\leq N+1\}$}{
 	  \vspace{-1mm}
 	  $\mathcal{P}= \mathcal{P} \cup \Big\{\left[x^{(i)}, SIR_i(x^{(i)})\right]\Big\} $ \vspace{-1mm}}
 	 }
 	 }
 	\hspace{-1mm}Derive\hspace{-.5mm} $x^{(n+N+1,n'+N+1)}_{\pm}\hspace{-1mm}$ and \hspace{-1mm} $x^{(n,n')}_{\pm}\hspace{-1mm}$, $\hspace{-.5mm}1\hspace{-.8mm}\leq \hspace{-.8mm}n<\hspace{-.8mm}n'\hspace{-.8mm}\leq \hspace{-.8mm}N+1\hspace{-.3mm}$, \hspace{-.8mm}using\hspace{-.5mm} Proposition\hspace{-.5mm}~\ref{prop:crosspointsSinlgeMulti}.\\
 	 \For{$ind \in \{+,-\}$}{
 	 \For{$(n,n')\in\{(n,n'): 1\leq n<n' \leq N+1\}$}{
 	 	 \If{\hspace{-.5mm}$SIR_n(x^{(n,n')}_{ind}) = \min\{SIR_{j}(x^{(n,n')}_{ind}):\hspace{-.5mm} 1\hspace{-.5mm}\leq\hspace{-.5mm} j\hspace{-.5mm}\leq\hspace{-.5mm} N\hspace{-.5mm}+\hspace{-.5mm}1\}$\hspace{-1.0mm} \textbf{and}\hspace{-.85mm} $SIR_n(\hspace{-.2mm}x^{\hspace{-.2mm}(\hspace{-.2mm}n,n'\hspace{-.2mm})\hspace{-.2mm}}_{ind}\hspace{-.2mm})\hspace{-.5mm} \geq\hspace{-.5mm} \min\{\hspace{-.4mm}SIR_{\hspace{-.2mm}N+1+j\hspace{-.2mm}}(\hspace{-.2mm}x^{\hspace{-.2mm}(\hspace{-.2mm}n,n'\hspace{-.2mm})}_{ind}\hspace{-.2mm})\hspace{-.5mm}:\hspace{-.5mm} 1\hspace{-.5mm}\leq\hspace{-.5mm} j\hspace{-.5mm}\leq \hspace{-.5mm}N\hspace{-.7mm}+\hspace{-.7mm}1\hspace{-.6mm}\}$}{
 	 \vspace{-.5mm}
 	  $\mathcal{P}= \mathcal{P} \cup \Big\{\left[x^{(n,n')}_{ind}, SIR_n(x^{(n,n')}_{ind})\right]\Big\} $}
 	  \vspace{-1mm}
 	 }
 	 	 \For{$(n,n')\in\{(n,n'): N+2\leq n<n' \leq 2N+2\}$}{
 	 	 	 \If{$\hspace{-.55mm}SIR_n (x^{\hspace{-.5mm}(n,n')\hspace{-.5mm}}_{ind})\hspace{-.5mm} = \hspace{-.5mm}\min\{\hspace{-.5mm}SIR_{N+1+j}(x^{\hspace{-.5mm}(n,n')}_{\hspace{-.5mm}ind}\hspace{-.5mm})\hspace{-.5mm}:\hspace{-.5mm} 1\hspace{-.5mm}\leq\hspace{-.5mm} j\hspace{-.5mm}\leq \hspace{-.5mm}N\hspace{-.65mm}+\hspace{-.65mm}1\hspace{-.5mm}\}$ \hspace{-.5mm}\textbf{and}\hspace{-.5mm} $SIR_n(x^{(n,n')}_{ind}) \hspace{-.5mm}\geq\hspace{-.5mm} \min\{SIR_j(x^{(n,n')}_{ind})\hspace{-.5mm}: \hspace{-.5mm}1\hspace{-.5mm}\leq\hspace{-.5mm} j\hspace{-.5mm}\leq\hspace{-.5mm} N+1\hspace{-.5mm}\}$}{
 	 \vspace{-.5mm}
 	  $\mathcal{P}= \mathcal{P} \cup \Big\{\left[x^{(n,n')}_{ind}, SIR_n(x^{(n,n')}_{ind})\right]\Big\} $}
 	  \vspace{-1mm}
 	 }
 	 }
 	 Derive $x^{(n,n'+N+1)}_{\pm}$, $1\leq n<n'\leq N+1$ using Proposition~\ref{prop:crosspointsDoubleMulti}.\\
 	 \For{$ind \in \{+,-\}$}{
 	 \For{$(i,j)\in \{(n,n'+N+1): 1\leq n<n'\leq N+1\}$}{
 	 \If{$\min\{SIR_j(x^{(i,j)}_{ind}): 1\leq j\leq N+1\} = SIR_i(x^{(i,j)}_{ind})$\hspace{-.5mm} \textbf{and} \hspace{-.5mm}$\min\{\hspace{-.5mm}SIR_{N+1+j}(\hspace{-.5mm}x^{(i,j)}_{\hspace{-.5mm}ind}\hspace{-.5mm})\hspace{-.5mm}: \hspace{-.5mm}1\hspace{-.5mm}\leq\hspace{-.5mm} j\hspace{-.5mm}\leq\hspace{-.5mm}N\hspace{-.5mm}+\hspace{-.5mm}1\hspace{-.5mm}\} = SIR_j(\hspace{-.5mm}x^{(i,j)\hspace{-.5mm}}_{\hspace{-.5mm}ind}\hspace{-.2mm})$}{
 	  $\mathcal{P}= \mathcal{P} \cup \Big\{\left[ x^{(i,j)}_{ind},SIR_i(x^{(i,j)}_{ind})\right]\Big\}  $
 	 }
 	 }}
 	 
 	 Consider $\mathcal{P}$ in the following format: $\mathcal{P}= \cup_{i=1}^{|\mathcal{P}|} \{\left[a_i,b_i \right]\}$\\
 	 $x^*_{\textrm{MSI}}=a_{i^*}$,  $i^*=\displaystyle\argmin_{i} \{b_i: \left[a_i,b_i \right]\in \mathcal{P},  -x^-_{jam}\leq a_i\leq x^+_{jam}  \}$.
 	  }
  \end{algorithm}
\noindent  the UAV; our method leads to considerably more (between $3.1 \textrm{dB}$ to $10.8 \textrm{dB}$) reduction in $\textrm{SIR}_{\textrm{max}}$. To better illustrate the performance gain, the percentage of reduction in $\textrm{SIR}_{\textrm{max}}$ upon using our method as compared to the baselines is depicted in Fig.~\ref{fig:2}, which reveals around $80\%$ (average) SIR reduction of our method.

Considering the jammer placement in the multi-hop setting with $D=5\textrm{km}$, we choose the position and the transmitting powers of UAVs randomly with respect to the following intervals: $x_i\in(0,5\textrm{km})$, $h_i\in [45,65]\textrm{m}$, $y_i\in [-10,10]\textrm{m}$, and $ p_i\in[20,25] \textrm{dBm}$, $1\leq i\leq N$. Each random assignment of the transmitting powers and positions of the UAVs is considered as one \textit{network realization}. Upon using the \textit{chasing a UAV} baseline, the jammer is placed underneath a randomly selected UAV in each network realization. Considering $20$ UAVs in the system, Fig.~\ref{fig:3} depicts $\textrm{SIR}_{\textrm{max}}$ for $20$ network realizations. As before, the best baseline method is chasing a UAV, which is considerably outperformed by our method. To reveal the performance gain, the average percentage of reduction in $\textrm{SIR}_{\textrm{max}}$ considering different numbers of UAVs in the network for $300$ network realizations upon using our method as compared to the baselines is depicted in Fig.~\ref{fig:4}, which shows a SIR reduction between $65\%$ to $97\%$ upon using our method. Examining Fig.~\ref{fig:4}, it is noteworthy to mention that as the number of UAVs increases, the performance gap between our method and the \textit{chasing a UAV} baseline decreases, which is vice versa considering the other two baselines. That is because, in general, considering a fixed distance between the TRs, as the number of UAVs increases and they get closer to each other, the position of the jammer becomes less important. Nevertheless, the \textit{chasing a UAV} baseline significantly deteriorates the SIR at only one UAV, which is the UAV located above the jammer. This makes this baseline method less effective as the number of UAVs increases since, considering \eqref{SIR_L_1_multi} and \eqref{SIR_L_2_multi}, there is a smaller chance that deteriorating the SIR at only a UAV corresponds to the decrease of both $\textrm{SIR}_{\textrm{Link\_1}}$ and $\textrm{SIR}_{\textrm{Link\_2}}$.
  \vspace{-3mm}
  \section{Conclusion}\label{conc}
  \vspace{-1mm}
\noindent
We proposed an effective approach for jammer placement in UAV-assisted wireless networks aiming to minimize the maximum achievable data rate of transmission of the system. We studied the problem for both the dual-hop and multi-hop relay settings. Given the non-convexity of the problem, we proposed a systematic tractable approach that can efficiently find the optimal placement of the jammer for both settings. As a future work, we suggest studying the problem when the UAVs can evade the interference by changing their locations. In this case, designing online adaptive algorithms for both the jammer and the UAVs is of particular interest.
\bibliographystyle{IEEEtran}
\bibliography{ABSbib}
\end{document}